\documentclass[prx,twocolumn,showpacs,floatfix,superscriptaddress,nofootinbib]{revtex4-1}

\usepackage{amssymb,amsmath,amstext}                
\usepackage{graphicx}                                               
\usepackage{epstopdf}                                               
\usepackage{color}                                                     
\usepackage{bm}                                                        
\usepackage{appendix}                                              
\usepackage[utf8]{inputenc}
\usepackage{bbold}
\usepackage{bbm}
\usepackage{ulem}
\usepackage{latexsym}
\usepackage[colorlinks=true,citecolor=blue,linkcolor=magenta]{hyperref}
\usepackage{standalone}
\usepackage{pgfplots}

\begin{document}

\title{Inhomogeneous quasi-adiabatic driving of quantum critical dynamics \\ in weakly disordered spin chains}

\author{Marek M. Rams}
\affiliation{Institute of Physics, Jagiellonian University,  \L{}ojasiewicza 11, 30-348 Krak\'ow, Poland }

\author{Masoud Mohseni}
\affiliation{Google Quantum AI, Venice, CA 90291, USA}

\author{Adolfo del Campo}
\affiliation{Department of Physics, University of Massachusetts, Boston, MA 02125, USA}

\begin{abstract}

We introduce an inhomogeneous protocol to drive a weakly disordered quantum spin chain quasi-adiabatically across a quantum phase transition and minimize the residual energy of the final state. The number of spins that simultaneously reach the critical point is controlled by the length scale in which the magnetic field is modulated, introducing an effective size that favors adiabatic dynamics.
The dependence of the residual energy on this length scale and the velocity at which the magnetic field  sweeps out the chain is shown to be nonmonotonic. We determine the conditions for an optimal suppression of the residual energy of the final state and show that inhomogeneous driving can outperform conventional adiabatic schemes based on homogeneous control fields by  several orders of magnitude.
\end{abstract}

\maketitle


\section{Introduction}
 
Techniques to control or assist adiabatic dynamics are of broad interest in quantum technologies, including quantum simulation and quantum computation \cite{DC08,CZ12}.
The breakdown of adiabatic dynamics in quantum critical systems is conveniently described using the  Kibble-Zurek mechanism (KZM) \cite{Kibble76,Zurek96,Dziarmaga10,Polkovnikov11,DZ14}.
This is the paradigmatic theory to account for the nonequilibrium dynamics across a continuous quantum phase transition. It exploits the divergence of the relaxation time in the neighborhood of the critical point in combination with scaling theory to predict the density of excitations in the final state that results from crossing the transition at a finite rate. 
The KZM can also be used to estimate the mean energy of the final state, known as residual energy, when measured with respect to the corresponding ground state energy \cite{DeGrandi10}.  Both the density of excitations and the residual energy are  shown to scale as a universal power law of the quench rate, where the power-law exponent is fixed by the correlation length and dynamic critical exponents, $\nu$ and $z$, as well as the dimensionality of the system. 

Strategies that have been developed to mimic adiabatic dynamics in quantum critical systems, in the absence of disorder, often boil down to finding ways out of the KZM. This is challenging as the KZM is broadly applicable and it holds even in strongly-coupled systems \cite{Chesler15,Sonner15}. Yet, a variety of protocols have been put forward. Prominent examples include the driving of finite many-body systems with a  nonzero energy gap \cite{Murg04}, coupling the system of interest to a thermal bath \cite{Patane08}, and engineering the time variation of the driving field using scaling theory \cite{BP08,SSM08} or optimal control \cite{OQC,PD13}.  Further approaches include the design of counterdiabatic fields to implement shortcuts to adiabaticity \cite{DRZ12,Takahashi13,Saberi14,Damski14,Campbell15}, or  the modulation of  multiple control parameters in time \cite{SS14}, see \cite{DS15} for a review. All these strategies are particularly crucial -- and should be additionally scrutinized -- in the presence of noisy control fields, that preclude adiabatic dynamics \cite{Dutta16}.

Recently, it has been shown that the KZM  should be modified when spatial or temporal inhomogeneities affect the critical behavior of classical and quantum systems. In the absence of disorder, the residual energy dependence on the quench rate can be enhanced in classical inhomogeneous systems \cite{KV97,Dziarmaga99,ZD08,Zurek09,ions10,DRP11,DKZ13}.  This is the case when the critical point is first reached at a local front that subsequently spreads throughout the system, completing the crossing of the phase transition. A number of recent experiments are consistent with the fact that the interplay between the velocity of the critical front and the speed of information 
way to suppressing defect formation \cite{DZ14}.  The role of causality has also been established in quantum spin chains with no disorder \cite{Dziarmaga2010a,Dziarmaga2010b,collura2010}.

By contrast, the development of techniques to approach the adiabatic limit in disordered systems is much more limited. This is however a pressing issue for boosting the performance of quantum annealers that encode combinatorial optimization problems and thus are inherently disordered  \cite{DC08}. It has been shown that, for disordered Ising spin chains, the KZM predictions are severely modified and the residual energy of the state upon completion of the driving exhibits only a weak dependence on the quench rate \cite{Dziarmaga2006,Caneva2007}: it no longer follows a power-law and vanishes only with the inverse of the logarithm of the quench rate. One main outstanding challenge, that we address in this work, is to explore the effect of inhomogeneous driving across a quantum critical point in the presence of disorder. The main goal is to spatially coordinate symmetry breaking events among neighboring regions by finding the appropriate degree of inhomogeneity and the speed of critical front to reduce the number of topological defects.

\begin{figure} [t]
\begin{center}
  \includegraphics[width=0.85 \columnwidth]{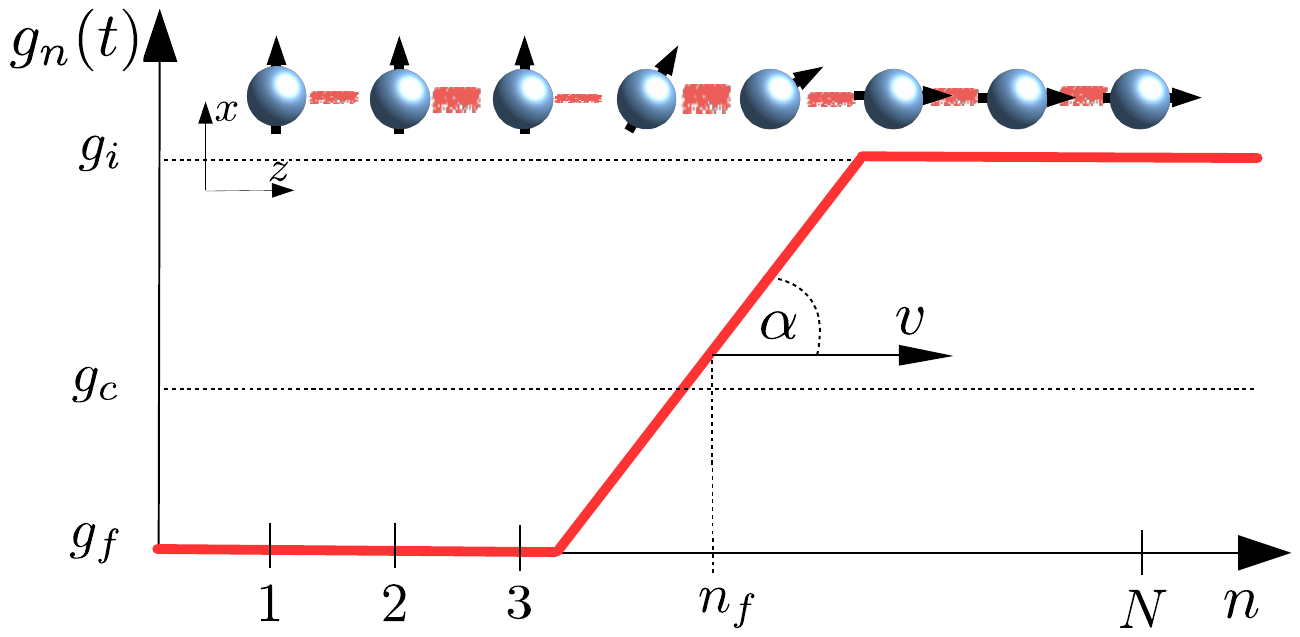}
\end{center}
  \caption{{\bf Driving  of a disordered spin chain by an inhomogeneous magnetic field. }  Under inhomogeneous driving, the critical front is reached locally as
 the  magnetic field is swept through the chain at  velocity $v$.   The length scale in which the external field is modulated, controlled by $\alpha$, sets the number of spins in the neighborhood of the critical point.
  } 
  \label{fig:scheme}
\end{figure}

In this work, we introduce an inhomogeneous driving strategy for a weakly disordered Ising spin chain by a transverse magnetic field with a smooth step-like spatial profile that   
sweeps out the chain from side to side, as illustrated in Fig.~\ref{fig:scheme}. The length scale in which the  field is modulated controls the number of spins that simultaneously cross the critical point. We study the dependence on the shape (slope) of the profile and the velocity at which  it is moved throughout the chain to minimize the residual energy of the final state, effectively approaching adiabatic dynamics.
We show that our local driving protocol can outperform conventional quantum annealing schedules based on homogeneous fields, reducing the relative residual energy by several orders of magnitude.

The paper is organized as follows: In Sec.~\ref{sec:2} we introduce the weakly disordered transverse Ising model and briefly review previous relevant results. In Sec.~\ref{sec:3} we use the adiabatic theorem to show the existence of a threshold velocity of the critical front below which the evolution is adiabatic. By an analysis in the spirit of the inhomogeneous KZM, we show that this velocity  has a universal character for smooth fronts. Subsequently, in Sec.~\ref{sec:4} we present simulations of full dynamics to find the inhomogeneous protocols that optimize the residual energy. While for fast driving a homogeneous control field proves advantageous,  in slower schemes for which  the critical front does not exceed the threshold velocity the inhomogeneous protocol with a smooth front extended over several spins turns out to be optimal, effectively reaching the adiabatic limit. We close with a discussion and conclusions in Sec.~\ref{sec:5}.


\section{Model}
\label{sec:2}
We consider a chain of $N$ spins described by the random Ising Hamiltonian 
\begin{equation} \label{eq:hamiltonian}
\hat H = - \mathcal{J} \sum_{n=1}^{N-1} J_n \sigma^x_n \sigma^x_{n+1} - \mathcal{J} \sum_{n=1}^N g_n \sigma^z_{n}  ,
\end{equation}
with quenched disorder (constant in time) in the nearest-neighbors couplings $J_n$.  
We set $\mathcal{J}=\hbar=1$ in  subsequent calculations, that  is equivalent to use $\hbar/\mathcal{J}$ as a unit of time.
Realizations of disorder are drawn from a uniform distribution
\begin{equation}
\label{eq:probability_density}
P(J_n) = \left\{ 
\begin{array}{cc}
1 &  \mathrm{for} ~ J_n \in (1/2,3/2),   \\
0 & \mathrm{otherwise}. 
\end{array} \right.
\end{equation}
The equilibrium properties of the model are well understood as it is solvable using the
strong disorder renormalization group approach \cite{Fisher1995}, see \cite{SDRGreview} for a review. In particular,  the critical point satisfies ${\log(g_c)} = \overline{\log(\vert J_n \vert)}$ for uniform $g_n = g_c$  ($\simeq 0.9558\ldots$), and belongs to the universality class of the infinite-randomess fixed point.

We consider a driving protocol where at an initial time $t_i$ the system is prepared in the ground state for the magnetic field $g_n(t_i)=g_i$ deeply in the paramagnetic phase (in the simulations we use $g_i = 3$) and then driven at a finite rate to the final value $g_n(t_f) = g_f$, deeply in the ferromagnetic phase. We set $g_f = 0$ for which the Hamiltonian in Eq.~\eqref{eq:hamiltonian} can be considered classical, with a  non-vanishing energy gap (notice that $J_n \ge 0.5 $  which we refer to as weak disorder) and is outside of the Griffiths phase surrounding the critical point.

Under homogeneous driving and in the absence of disorder the nonequilibrium dynamics is well described by the KZM \cite{Kibble76,Zurek96,Dziarmaga10,Polkovnikov11,DZ14}. The mechanism exploits the divergence of the equilibrium relaxation time $\tau[\varepsilon]=\tau_0/|\varepsilon|^{z\nu}$, known as critical slowing down, as a function of the dimensionless distance  to the critical point  $\varepsilon=(g_c-g)/g_c$. In the proximity of $g_c$ the modulation of the magnetic field can be linearized in the form $g(t)=g_c(1-t/\tau_Q)$,  so that $\varepsilon=t/\tau_Q$, where $\tau_Q$ sets the quench time. The critical point is reached at $t=0$ around which the relaxation time diverges. As the system is driven through the phase transition, the evolution can be roughly split in three sequential stages where the dynamics is approximately adiabatic, frozen and adiabatic again. The time scale in which  the system leaves the frozen stage to evolve adiabatically in the broken-symmetry side of the transition is known as the freeze-out time $\hat{t}=(\tau_0\tau_Q^{z\nu})^{\frac{1}{1+z\nu}}$, and satisfies $\tau(\hat{t})=|\varepsilon/\dot{\varepsilon}|$.  KZM estimates the size of the domains in the broken symmetry phase using the equilibrium value of the correlation length $\xi[\varepsilon]=\xi_0/|\varepsilon|^{\nu}$. 
At the freeze-out time, it scales as a power-law of the quench rate, e.g., $\hat{\xi}=\xi[\varepsilon(\hat{t})]=\xi_0(\tau_Q/\tau_0)^{\frac{\nu}{1+z\nu}}$. In the Ising model without disorder, $\nu=z=1$, 
and the KZM prediction for the density of excitations reads $d\sim \xi(\hat{t})^{-1} \sim \frac{1}{\sqrt{\tau_Q}}$ \cite{Dziarmaga2005,Dorner2005,Polkovnikov2005},
as corroborated by the exact solution \cite{Dziarmaga2005}. Similar power-laws  can be derived for other observables using scaling theory \cite{DeGrandi10,Kolodrubetz12,Francuz16}. This is the case for the residual mean energy defined as the difference between the mean energy of the system following the completion of the protocol and the corresponding ground state energy, e.g., $Q=\langle \hat{H}\rangle_\tau-E_{\rm gs}$. For the Ising model,   the value of $Q$ after the quench scales in the same way as the density of excitations. 

The presence of disorder, that drives the universality class of the critical point towards the infinite-randomnes fixed point, severely modifies the dependence of the relaxation time on the distance from the critical point. It is found that $\tau[\varepsilon] \simeq \tau_0/|\varepsilon|^{1/|\varepsilon|}$, where the critical exponent $z=1/|2\varepsilon| + O(1)$ effectively diverges as the system approaches the critical point ($z \to \infty$  as $\varepsilon \to 0$) \cite{Fisher1995}. 
As a result, the scaling of the density of excitations is no longer described by a power-law and a more careful analysis based on KZM predicts $d\sim1/\ln^2\tau_Q$  in a slow transition (for large $\tau_Q$) \cite{Dziarmaga2006,Caneva2007}.  The dependence on the quench time becomes then much weaker than in the absence of disorder. We note that a similar result can also arise in a clean model as a result of a particular decoherence mechanism \cite{Cincio2009}.

The existence of these logarithmic scaling laws signifies that one is forced to consider exponentially long evolution times to reduce the residual energy of the final state.
As an alternative to a  global homogeneous modulation of the magnetic field, we introduce an inhomogeneous protocol
\begin{equation} \label{gn_inhomo}
g_n(t) = \left\{ 
\begin{array}{ll}
g_i, & n- v t > \frac{g_i-g_f}{2 \alpha},   \\
\frac{g_i+g_f}{2} + \alpha (n-v t), & |n- v t| \le \frac{g_i-g_f}{2 \alpha},    \\
g_f  &n- v t < \frac{g_f-g_i}{2 \alpha},
\end{array} \right.
\end{equation}
where the linear front interpolating between $g_i$ and $g_f$ travels through the chain with velocity $v$ and gradually drives the system from the paramagnetic to the ferromagnetic phase, by sweeping out the chain from end to end. We denote by  $n_f = v t$ the position of the spin for which the magnetic field equals the arithmetic mean of $g_i$ and $g_f$. The slope $\alpha$ sets the effective number of spins  driven at a given instant.  The resulting protocol is illustrated in Fig.~\ref{fig:scheme} and interpolates between the following two limiting cases: (i) homogeneous driving, which is recovered in the limit of $\alpha \to 0$ and $v \to \infty$ while keeping $\alpha v = \tau_Q^{-1}$ fixed, and (ii) driving of one spin at a time, when $\alpha \approx 1$ \cite{Schaller2008}.

In absence of disorder, the KZM can be extended to account for an inhomogeneous scenario \cite{ZD08,Zurek09,ions10,DRP11,DKZ13}. The central prediction is the existence of threshold velocity $v_t$ that determines the relevance of the driving scheme. When the velocity of the front  widely surpasses this threshold value, $v \gg v_t$,  the effect of the local modulation of the control magnetic field is negligible and the nonadiabatic critical dynamics resembles  that under homogeneous driving, well described by the standard KZM.  By contrast,  in the case  $v \ll v_t$  the length scale in which the front is smoothed out becomes relevant and determines the number of spins that simultaneously experience criticality. The smooth  front of the inhomogeneous field opens up an energy gap, that allows for adiabatic evolution.

For a smooth front with $\alpha \ll 1$, KZM predicts that the penetration depth across the critical point, i.e., 
the size of the critical region separating the phases to both sides of the inhomogeneous front,  varies as \cite{ZD08,platini2007,collura2009} 
\begin{equation}
\hat \xi_i \sim \alpha ^{-\frac{\nu}{\nu+1}}.
\label{eq:IKZM_xi}
\end{equation}
 When the gap at the critical point vanishes polynomially with the system size
this leads to the opening of instantaneous gap that scales as $\hat \Delta_i \sim \hat \xi_i^{-z}  \sim \alpha ^{\frac{z \nu}{\nu+1}}$.
By combining the characteristic time  and length scales in the problem one can then estimate the threshold velocity as
\begin{equation} \label{vc_KZ}
v_t \sim \alpha ^{\frac{(z-1) \nu} {\nu +1}}.
\end{equation}
In the Ising model without disorder $z=1$  and $v_t$ is a constant independent of $\alpha$ ($\alpha \ll 1$). The analytical solution \cite{Dziarmaga2010a} shows that when $J_n =1$ in Eq.~\eqref{eq:hamiltonian}, $v_t=2$ and equals the sound velocity at the critical point.  What is more, the transition between the adiabatic regime for $v < v_t$ and the effectively homogeneous regime for $v > v_t$ is actually sharp.

The presence of disorder changes the universality class of the model and the assumptions leading to Eq.~\eqref{vc_KZ} do not longer hold. Naively setting $z \to \infty$ in that equation leads to a vanishing  threshold velocity and would indicate  that inhomogeneous driving does not favor adiabatic dynamics in disordered systems.  In this article we show this not to be the case. Indeed, the analysis presented in the next Section predicts that  the threshold velocity $v_t$ acquires a finite value, that admits a universal form for small $\alpha$. This paves the way to implement adiabatic dynamics by inhomogeneous driving. 

All simulations shown below are done using the Jordan-Wigner transformation that maps the Hamiltonian in Eq.~\eqref{eq:hamiltonian} onto the system of free-fermions where it can be solved numerically in a standard way. For details, we refer the reader to Appendix B.

\section{Threshold velocity at the adiabatic limit}
\label{sec:3}
We next provide a quantitative prediction of the threshold velocity under quasi-adiabatic dynamics when diabatic transitions  occur  within the manifold spanned by the ground and the first-excited states. The formation of excitations is proportional to the  mixing matrix element between these two states,
\begin{equation}
 |\langle 0,t | \frac{d\hat H}{dt} |1,t \rangle| = |\langle 0,t | \frac{d \hat H}{dn_f} |1,t \rangle| \frac{dn_f}{dt} = \Omega(n_f) v,
\end{equation}
where we have defined $\Omega(n_f) = |\langle 0,t | \frac{d \hat H}{dn_f} |1,t \rangle| $. The instantaneous energy gap $\Delta_{n_f}$ can be parameterized both as a function of the front position $n_f$ and the time of evolution $t$,
\begin{equation}
\Delta_{n_f}=\Delta(n_f)=\Delta(t) = E_1(t) - E_0(t),
\end{equation}
where $E_{0,1}(t)$ are the energies of the instantaneous ground state $ |0,t \rangle$ and the first excited state $ |1,t \rangle$ of Hamiltonian \eqref{eq:hamiltonian}, respectively. Since the parity operator $\hat P = \prod_{i=1}^N \sigma^z_n$ is conserved during time evolution, we consider only the subspace with the same parity as the initial ground state.

According to the adiabatic theorem, the evolution follows the instantaneous ground state with high fidelity provided that 
\begin{equation}
\label{eq:AT}
\frac{|\langle d\hat H/dt \rangle|_{0,1} }{\Delta(t)^2} = \frac{ v \Omega(n_f)} {\Delta(n_f)^2} \ll 1,
\end{equation}
which allow us  to estimate  the value of threshold sweep velocity $v_t$ below which the dynamics of the quench is effectively adiabatic as 
\begin{equation}
\label{eq:local_vc}
v  \ll v_{n_f} = v_t(n_f) = \frac{\Delta(n_f)^2}{4 \Omega(n_f)}.
\end{equation}
The factor of $4$ in the definition of $v_t$  above is introduced so that $v_t(n_f)$ matches the exact known value in the case without disorder, when $v_t=2$, see Appendix A for a detail discussion of this case. 

The matrix element reads
\begin{equation}
\Omega(n_f) = |\langle0,n_f| \sum_n g'_n(n_f) \sigma_n^z  |1,n_f\rangle|, 
\label{eq:Omega_nf}
\end{equation}
where 
\begin{equation}
g'_n(n_f) = \left\{ 
\begin{array}{ll}
0, & |n- n_f| > \frac{g_i-g_f}{2 \alpha},   \\
-\alpha, & |n- n_f| \le \frac{g_i-g_f}{2 \alpha}.  
\end{array} \right.
\end{equation}

This analysis is restricted to quasi-adiabatic dynamics governed by adiabatic following of the ground-state and $|0\rangle$-$|1\rangle$ transitions. As a result, it is expected to fail when transitions to higher excited states are dominant, e.g.,  for large $\alpha$ ($\approx1$), when the inhomogeneous front approaches a step function.

\begin{figure} [t]
\begin{center}
\includegraphics[width=0.90 \columnwidth]{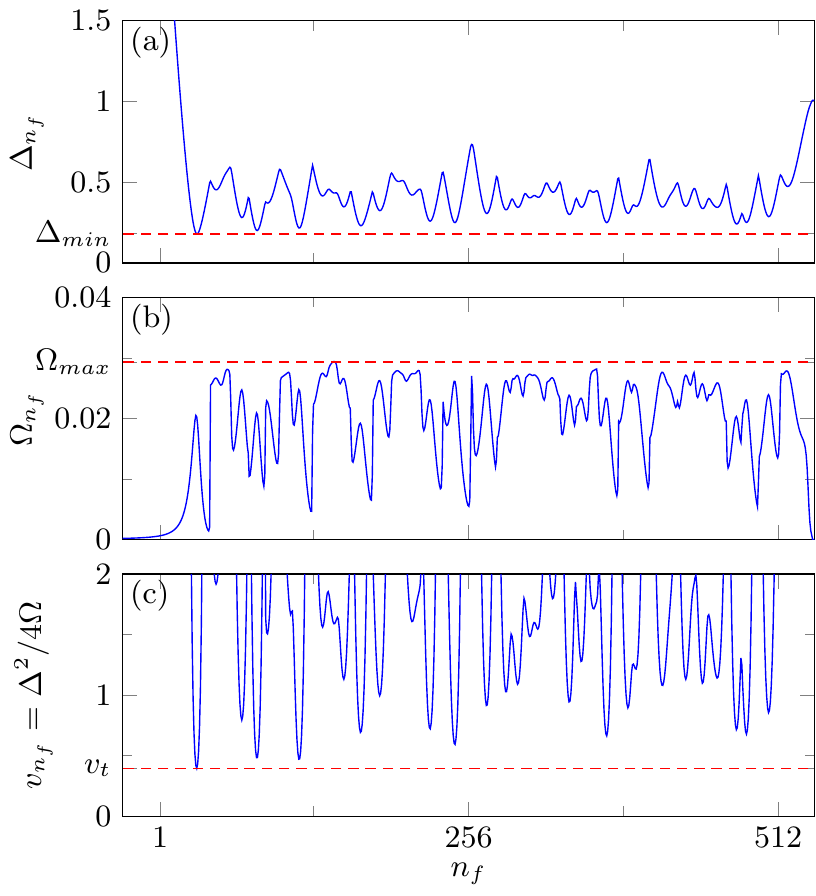}
\end{center}
  \caption{{\bf Local adiabaticity under inhomogeneous driving.} Instantaneous (a) gap $\Delta(n_f)$ and (b) mixing term $\Omega(n_f)$ when the inhomogeneous front, centered at site $n_f$, as it travels through the chain for a given realization of disorder. 
   (c) From this trajectory, a local  value is estimated for the velocity below which the system should stay adiabatic  
  and the global minimum is identified as  threshold velocity  ($N=512$, $\alpha = 1/32$).  } 
  \label{fig:2}
\end{figure}

\begin{figure} [t]
\begin{center}
\includegraphics[width=0.95\columnwidth]{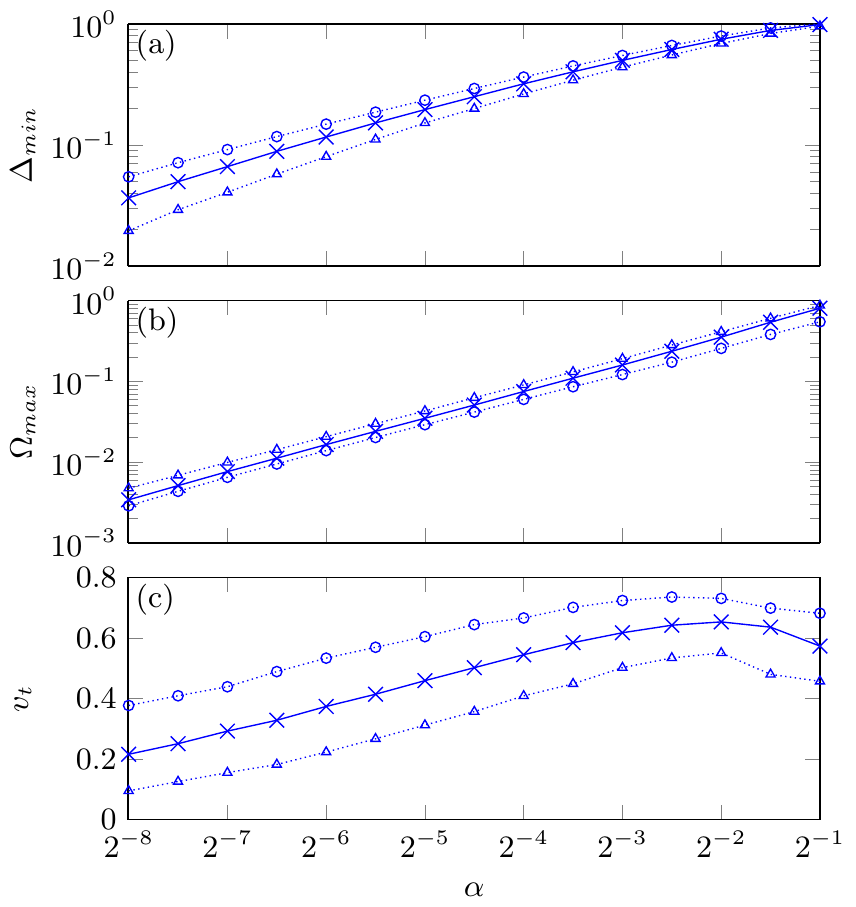}
\end{center}
  \caption{ 
{\bf Local adiabaticity as a function of the front shape of the inhomogeneous control field.} Dependence of (a) minimal instantaneous gap, (b) maximal mixing term, and (c) threshold velocity 
as a function of the slope $\alpha$ of the  magnetic field. A solid line denotes the median of 1000 realizations while dotted lines show $5\%$ (circle, easiest instances) and $95 \%$ (triangles, hardest instances) quantiles, respectively. For each realization $\Delta_{min}$, $\Omega_{max}$ and $v_t$ are extracted as indicated in Fig.~\ref{fig:2} ($N=512$).} 
  \label{fig:3}
\end{figure}

In Figure \ref{fig:2}(a) we show the instantaneous gap during the evolution for a single realization of disorder and fixed value of the slope $\alpha=1/32$. While the gap  fluctuates as the front travels through the chain, it remains finite. We define $\Delta_{min}$ as the minimal gap for a given realization of disorder.  This definition involves averaging over a finite-length chain ($N=512$ in Figs.~\ref{fig:2} and \ref{fig:3}) and in principle depends on $N$. However, since only a fixed fraction of the system is being driven at given instant, we expect the dependence to be weak for  finite $\alpha> 0$. In that case,
the critical front can be thought of as probing different {\it local} realizations of disorder where the effective size of the system $\hat \xi_i$ (set by $\alpha$) is finite. As a result, fluctuations of the instantaneous gap are limited and there is a negligible probability of having a gap smaller than a given threshold.  This can be qualitatively seen in Fig.~\ref{fig:3}(a) where we consider many realizations of disorder and show different quantiles of $\Delta_{min}$, as  we shall discuss  further at the end of this section where we provide a quantitative scaling argument.
By contrast,  in the homogeneous case ($\alpha=0$)  the typical energy gap at the critical point is expected to vanish as $\sim \exp(- C\sqrt{N})$ \cite{Fisher1995,Young1996,Fisher1998,Caneva2007}, where $C$ is a nonuniversal constant.

In Fig.~\ref{fig:2}(b) we show the mixing term $\Omega(n_f)$ for the same realization of the couplings, and by combining it with the gap, from Eq.~\eqref{eq:local_vc} we estimate the local value of threshold velocity of the front $v_t(n_f)$ below which the evolution is expected to stay adiabatic. We define $v_t$ as the minimum of $v_t(n_f)$ for a given realization; see Fig.~\ref{fig:2}(c), where its value is of order unity for  $\alpha=1/32$. Since $v_t(n_f)$ is widely fluctuating one could envision a fine-tuned inhomogeneous driving protocol where the velocity of the front is adjusted with respect to the local value of the threshold velocity. However, the design of the corresponding driving protocol $g_n(t)$ would require quite a specific knowledge of the system.  Here, to explore the broad applicability and universality of the inhomogeneous scheme, we keep the front velocity $v$ fixed along the evolution.

 The results obtained from sampling over many realizations of disorder are summarized in Fig.~\ref{fig:3}. It shows the dependence of the  average value of the minimum gap $\Delta_{ min}$, the largest value of mixing matrix denoted by $\Omega_{ max}$ as well as the threshold velocity $v_t$ as a function of the smoothness $\alpha$ of the inhomogeneous magnetic-field front for a disordered Ising chain. The average is taken over 1000 realizations from which statistics is built to  determine the quantiles corresponding to the hardest ($95\%$) and easiest  ($5\%$) cases, displayed by dotted lines.
All quantities  increase  monotonically as a function of $\alpha$ in the range of values considered, with the exception of the threshold velocity that levels off for large values of $\alpha$.

For smooth fronts corresponding to small values of $\alpha$, a power-law fit yields  $\Omega_{max} \sim \alpha^{1.1}$  similar to the Ising case with no disorder where $\Omega \simeq \alpha$. The gap $\Delta_{min}$, however, disappears faster than polynomially in that limit which results in a monotonic dependence of $v_t$ on $\alpha$ (without disorder $\Delta_{min} \sim \alpha^{1/2}$). The maximum value of velocity is obtained for a value of $\alpha$ close to but lower than  unity, when the inhomogeneous front extends over a few sites. We also notice that the optimal $\alpha$ is smaller than $1$ (i.e. the limit of driving one spin at a time).

\begin{figure} [t]
\begin{center}
\includegraphics[width=0.95\columnwidth]{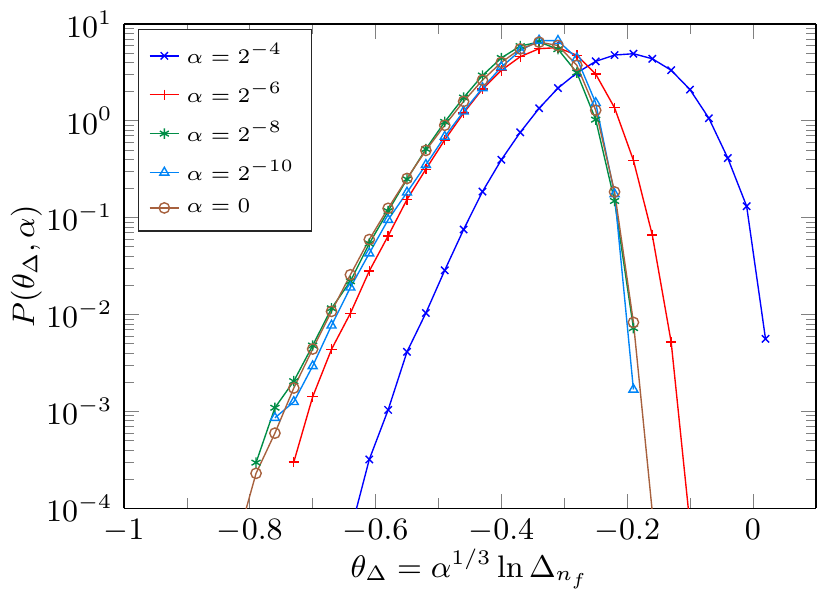}
\end{center}
  \caption{{\bf Universal gap distribution for different fronts of the inhomogeneous control field.}  
  The distribution of  instantaneous gap $\Delta_{n_f}$ is found when the front is traveling inside the chain.  The distributions for different $\alpha \ll 1$ collapse onto each other when the scaling variable $\theta_\Delta$ is used. For the distribution of the gap with a homogeneous magnetic field at the critical point ($\alpha=0$),  the scaling variable reads $\theta_\Delta = \log(\Delta)/\sqrt{0.46 N}$ \cite{comment_gaps}.}
  \label{fig:4}
\end{figure}
 
For a more quantitative scaling prediction, we consider the distribution of the instantaneous gap $\Delta(n_f)$, where the front is traveling inside the chain as in the middle part of Fig.~\ref{fig:2}(a).
 In doing so, we disregard the configurations in  which the front is entering or leaving the chain and the gap is large. We denote by $P(\Delta_{n_f},\alpha)$ the distribution of the instantaneous gap for different $\alpha$, normalized according to $\int_0^{\infty} d \Delta_{n_f} P(\Delta_{n_f},\alpha) = 1$.
For  homogenous driving ($\alpha=0$) of a finite system at the critical point the equivalent distribution over realizations of disorder is universal if one considers the scaling variable $\theta = N^{-1/2} \ln \Delta$ \cite{Fisher1995,Young1996,Fisher1998,Caneva2007}.
In the spirit of the inhomogeneous KZM \cite{DKZ13}, for an inhomogeneous system the characteristic length scale which governs the behavior of the system is expected to scale as $\hat \xi_i \sim \alpha^{-\frac{\nu}{1+\nu}} \sim  \alpha^{-2/3}$, see Eq.~\eqref{eq:IKZM_xi}. This suggests that the relevant scaling variable for small finite $\alpha$ is 
\begin{equation}
\label{eq:theta_delta}
\theta_\Delta = \hat \xi_i^{-1/2} \ln \Delta_{n_f} = \alpha^{1/3} \ln \Delta_{n_f},
\end{equation}
 where $\hat \xi_i$ plays the role of an effective size of the critical system.  For the random Ising model, the critical exponent $\nu=2$ \cite{Fisher1995} characterizes the equilibrium value of  $\xi[\varepsilon]$, that describes mean correlations dominated by rare pairs of strongly correlated spins and should be relevant for the low energy part of the spectrum.

\begin{figure*}[t]
    \includegraphics[width=\linewidth]{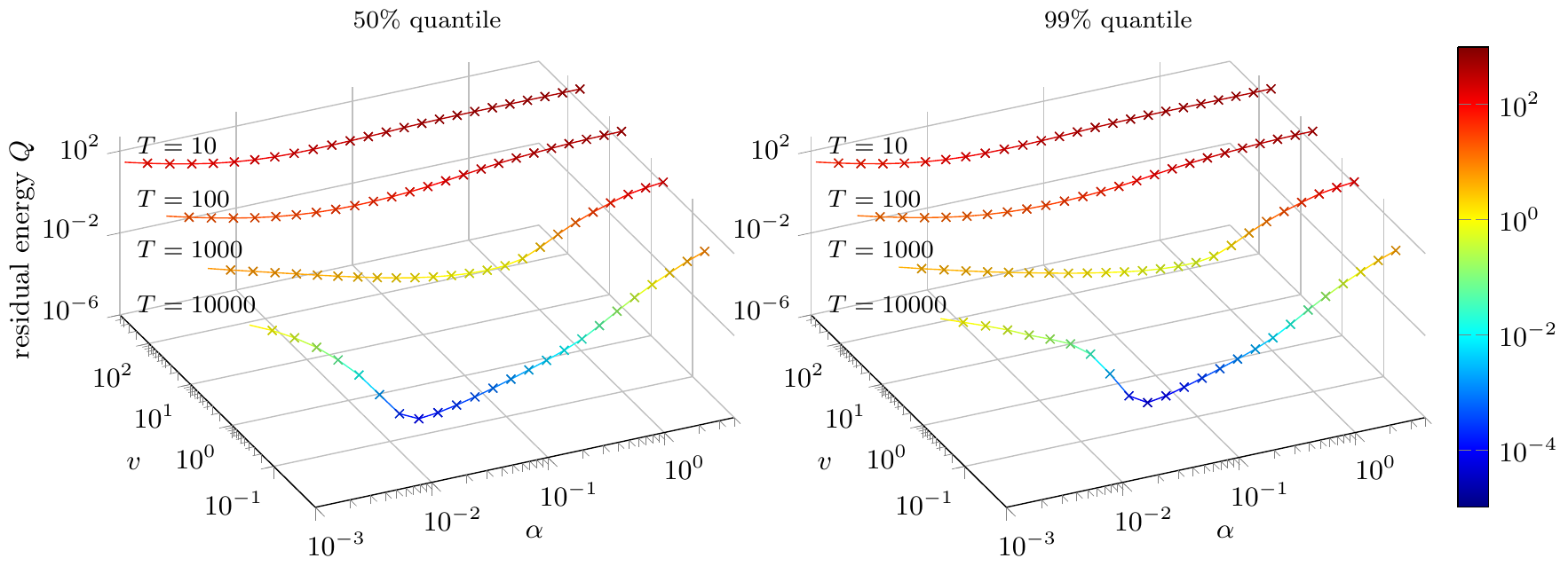} 
  \caption{{\bf Optimization of the  inhomogeneous driving protocol. } Landscape plots of residual energy for various time scales and $\alpha(v)$ for  $N=512$.
  We show the $q=50 \% $  and $q=99 \%$ quantiles ($q$ is the percentage of realizations with smaller residual energy) obtained from simulation of $1000$ realizations ($500$ for $T=10^4$). While for short time scales the best residual energy is obtained for homogeneous driving, for longer time scales the sharp minimum appears for intermediate values $1/16 \ge \alpha \ge 1/32$. The residual energy for optimal smoothness, $\alpha\simeq0.03$  at $T=10000$ timescale, is five orders of magnitude smaller than both the standard annealing strategy with $\alpha=0$ and the inhomogeneous scheme based on flipping one spin at a time, shown for $\alpha=2\sqrt2\simeq2.82$ \cite{comment_discret_alpha}. } 
  \label{fig:5}
\end{figure*}

In order to numerically verify Eq.~\eqref{eq:theta_delta}, for each realization of disorder we sample  values of $\Delta_{n_f}$ for a couple of hundreds equidistant instances of $n_f$ corresponding to the front traveling inside the chain; see. Fig~\ref{fig:2}(a). We  collect the values of the gap obtained that way for a couple of thousands realizations and
from the histogram we extract the probability distribution $P(\theta_\Delta,\alpha)$ as  a function of $\alpha$. Depending on $\alpha$ the statistics is built from $>10^6$ points from $>2000$ realizations of disorder. The distributions for different $\alpha \ll 1$ collapse onto each other corroborating our scaling prediction;  as seen for  $\alpha \le 1/64$ in Fig.~\ref{fig:4}. In addition, this distribution coincides with the one obtained for homogeneous system at the critical point ($g_n = g_c$), when the scaling variable $\log(\Delta)/\sqrt{0.46 N}$ is used \cite{comment_gaps}. The factor of $\simeq0.46$, which we found by collapsing the distributions in Fig.~\ref{fig:4}, can be interpreted as the prefactor  in the scaling $\hat \xi_i \sim \alpha^{-2/3}$, that sets the effective size of the critical region.

On the other hand we observe that for larger values of $\alpha$, such as e.g.~$\alpha=2^{-4}$ in Fig.~\ref{fig:4}, the distribution differs from the universal one. We expect that this happens due to the effective finite-size effect, where the size of the critical region is so small that non-universal contributions are still relevant.  While the analytical strong disorder renormalisation group approach of Ref. \cite{Fisher1995,SDRGreview} cannot be directly used in the presence of  inhomogeneous (position correlated) field,  we  expect that a number of initial decimation RG steps would be necessary to approach universal fixed point trajectory. 

The typical value of instantaneous gap scales  as $\Delta_{n_f} \sim \exp(-C \alpha^{-1/3})$ in the limit $\alpha \to 0$. Here, we are  mostly interested in   the minimal gap $\Delta_{min}$,  which would be determined by the behavior of the tail of $P(\theta_\Delta,\alpha)$ corresponding to small energies. The derivation in \cite{Fisher1998} suggests that we can expect this tail to be Gaussian,  which is indeed consistent with the data in Fig.~\ref{fig:4} \cite{comment_gaps}. As the front travels though the chain, it samples the distribution $P(\theta_\Delta,\alpha)$ in a continuous way; see Fig.~\ref{fig:2}(a).  We can  estimate the dependence of the minimal gap $\Delta_{min}$  on the system size $N$ by assuming that, to leading order, $N$ instances are drawn from the distribution $P(\theta_\Delta,\alpha)$  and determining the probability distribution for the minimal value. With a Gaussian tail this means that any fixed quantile of the minimal (global) gap (e.g. plotted in Fig.~\ref{fig:3}(a) for $N=512$), vanishes  slower than a  polynomial in $N$, making this dependence a subleading correction. The leading contribution related to the system size is the time needed for the front to travel though the chain with fixed velocity which is proportional to $N$  (for fixed $\alpha$).

Summarizing, the above analysis suggests the existence of a finite threshold velocity $v_t$ for non-zero $\alpha$ and a maximum at $\alpha$ near unity, when the front extends over few sites. However, given our analysis  in terms of the the instantaneous ground state and the first excited state, the values of $v_t$ could in principle  be overestimated. This is especially true for large $\alpha$ close to unity, when the first excited state is not well separated from the rest of the spectrum. This could be addressed by using adiabatic theorem taking into account all excited states, see e.g. \cite{Kato50,Comparat2009}. We however take a different approach in the next section,  namely, the numerically-exact simulation of the full dynamics.

\section{Optimization of the protocol and residual energy}
\label{sec:4}
Given the existence of a finite threshold velocity $v_t$ discussed in the previous section,  we next explore the possibility of implementing adiabatic dynamics under inhomogeneous driving. 
In particular, we focus on the  minimization of the residual energy $Q$ of the final state. 

We note that the total time required for the inhomogeneous protocol to be completed reads  
\begin{equation}
T = \frac{N}{v}+ \frac{|g_i-g_f |}{\alpha v},
\label{InhomoT}
\end{equation}
where the first term corresponds to the time needed for the middle of the front to travel through the chain and the second term accounts for additional time needed for the magnetic field to reach  the final value for all the spins.
In the homogeneous limit, the second term in (\ref{InhomoT}) dominates and the total time reduces to  $|g_i-g_f | \tau_Q$ (in this limit $v \to \infty$ as  $\alpha \to 0$).
By contrast, in the strongly inhomogeneous limit (when $N \gg |g_i-g_f |/\alpha$) the first term dominates and the second one constitutes a small over-head.
Here, to compare different protocols we fix the value of $T$ and choose the velocity $v$ according to Eq.~\eqref{InhomoT}  for a given $\alpha$ and $N$.

A direct  analysis of the performance of the inhomogeneous driving scheme is shown in Figure  \ref{fig:5} that displays the dependence of $Q$ on both $v$ and $\alpha$ for a fixed value of the ramp time $T$. The slope $\alpha$ in this plot  interpolates between the nearly homogeneous transition and the limit of a steep front where only one spin is driven at a time. We observe that for short time scales the homogeneous driving is optimal. However, for longer ramps, when the velocity of the front is small enough, there is a sharp minimum for intermediate values of $\alpha$ where the dynamics is effectively adiabatic. The value of the threshold velocity that marks the appearance of that minimum is consistent with $v_t$, obtained in Fig.~\ref{fig:3} for intermediate and small values of $\alpha$. Notice, however, that the actual minimum of $Q$ is reached for a slightly smaller value of $\alpha$ than suggested by that figure. 

This efficiency in suppressing excitations is shown to be  fairly insensitive to the hardness of the problem as quantified by the quantiles considered. Approximately the same landscape is observed for quantiles  with  $q=50\%$ and $q=99\%$, where  $q$ marks the percentage of realizations which have smaller residual energy.  Further,  the optimal value of $\alpha$ 
weakly depends on the quantile. In particular,  for $T=10^4$, $\alpha \approx 1/16$ is optimal for harder cases with $q=99\%$ and $\alpha \approx 1/32$ is optimal for  $q=50\%$ \cite{comment_discret_alpha}. Notice however that in both cases the value of the residual energy is almost that of an adiabatic transition, with the fidelity larger than $0.9999$ in both cases. In this limit the actual position of $Q$  might also depend on the additional smoothing of the critical front in Eq.~\eqref{gn_inhomo} that is nonanalytic at the point between the piecewise linear and constant sections of the front \cite{Lidar2009}. 

\begin{figure}[t]
    \includegraphics[width=\linewidth]{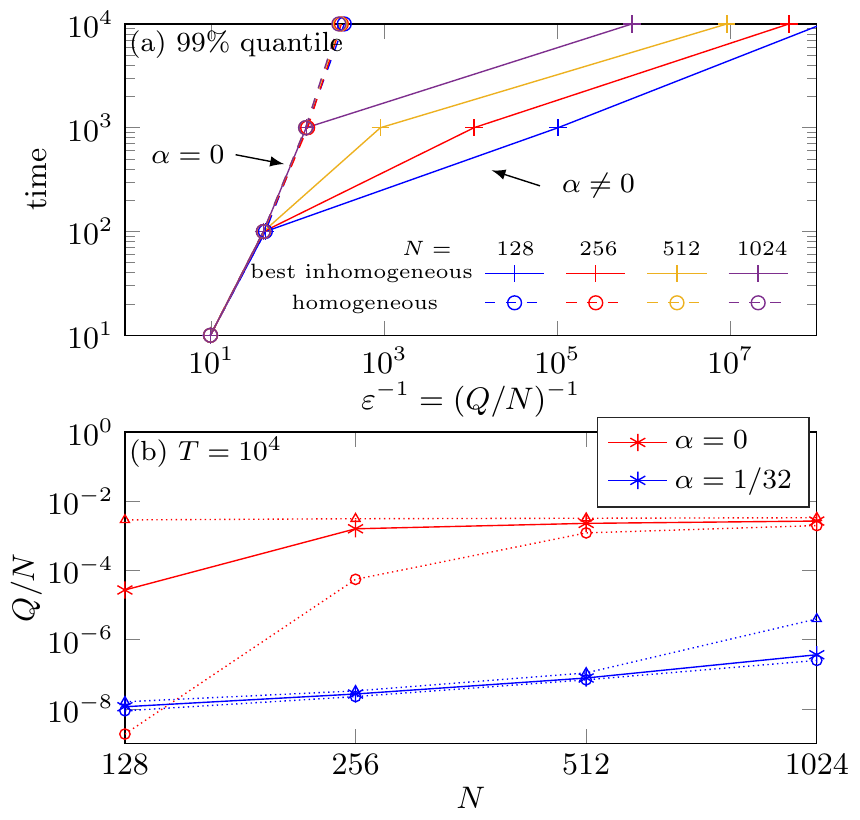} 
  \caption{{\bf Supremacy of optimal inhomogeneous driving over homogeneous schemes.} 
  Comparison of the homogeneous and best inhomogeneous protocol for different ramp times and system sizes. (a) $99\%$ quantile of residual energy. The symbol $+$ denotes the  best inhomogeneous strategy, where the optimal $\alpha$ is extracted from landscape plots in Fig.~\ref{fig:5}. For long-enough timescales it is advantageous to use larger $\alpha\approx 1/16$, while for short ramps homogeneous driving $\alpha \to 0$ is better. Circles show the residual energy for a homogeneous quench in the same ramp time. (b) Comparison of  homogeneous (red) and inhomogeneous (blue) strategies with fixed $\alpha = 1/32$ ($T= 10000$). Solid lines indicate the median of $1000$ realizations for $4$ system sizes while the dotted lines mark $1\%$ and $99\%$ quantiles, respectively. } 
  \label{fig:6}
\end{figure}

Finally, we identify scenarios for the supremacy of the inhomogeneous driving scheme over its homogeneous counterpart  in Fig.~\ref{fig:6}. In particular,  we plot in Fig.~\ref{fig:6}(a) the time needed to reach a given quality of the solution, as quantified  by the inverse of residual energy density. We compare the two schemes, when the optimal value of $\alpha$ for the inhomogeneous scheme is found from the landscape studies as in Fig.~\ref{fig:5}. 
The homogeneous transition (or $\alpha \to 0$) is shown to be optimal for short time scales in Fig.~\ref{fig:6}(a). However, for long time scales the residual energy after such a quench is expected to scale only as $Q/N  \sim 1/\log( T)^{2}$ \cite{Dziarmaga2006,Caneva2007}, making it unpractical to reach the adiabatic limit. The inhomogeneous driving approaches this limit for non-zero $\alpha$ for long enough time scales such that the velocity of the inhomogeneous front is  reduced below the threshold value. 
This is further confirmed in Fig.~\ref{fig:6}(b) where we compare the performance of homogeneous and inhomogeneous protocols with fixed $\alpha=1/32$ for a time-scale of $T=10^4$, for different system sizes. As  the system size is reduced  for given $T$, the velocity of the front is proportionally smaller, which is the reason behind the weak increase of residual energy with growing $N$ in the inhomogeneous case.

\section{Conclusions}
 \label{sec:5}
 
In summary, we have analyzed the driving of weakly-disordered spin chains with a time-dependent magnetic field. Under spatially homogeneous driving, the minimization of the residual energy in the final state is essentially constrained by the adiabatic theorem. Long-evolution time are then required to minimize excitations. As an alternative scheme, we have proposed the use of an inhomogeneous magnetic field  that sweeps out the system at a well-controlled velocity. In this scenario the spatial modulation of the magnetic field introduces an effective system size that favors adiabatic dynamics. The dependence of the residual energy of the final state on the latter and the sweeping velocity is not monotonic.
Upon optimization with respect to these two parameters, we have identified the supremacy of  inhomogeneous driving over homogeneous schemes in reducing the residual energy of the final state.
In this article we have focused on the case of weak-disorder where the couplings $J_n$ are nonzero. We shall discuss the case of strong disorder  with possibly vanishing couplings $J_n$
in a subsequent article \cite{MF_inprep}.

Our results can prove useful in the design of novel quantum annealing protocols with inhomogeneous control fields on disordered spin systems with the potential to outperform conventional schemes \cite{QuantumAnnealing} and might be applicable to open systems \cite{Vadim2015}. 
Inhomogeneous schedules with controllable local magnetic field can be implemented on the next generations of quantum annealers that are currently under development by D-Wave System and the Google Quantum AI laboratory. Our approach might be applied for finding higher quality of solutions for constrained optimization problems over standard adiabatic quantum computation (with homogeneous fields), as the corresponding embedded problems on annealing hardware Hamiltonians would involve finding low-energy states of disordered spin glasses on low-dimensional lattices.

\section{Acknowledgments}
We acknowledge funding support from UMass Boston under project P20150000029279 (A.d.C.) and Narodowe Centrum Nauki (NCN, National Science Center) under Project No. 2013/09/B/ST3/01603 (M.M.R), and Quantum Artificial Intelligence Laboratory at Google (M.M.) 

\appendix



\section{Adiabatic theory approach to clean Ising model}
\begin{figure}[b]
    \includegraphics[width=\linewidth]{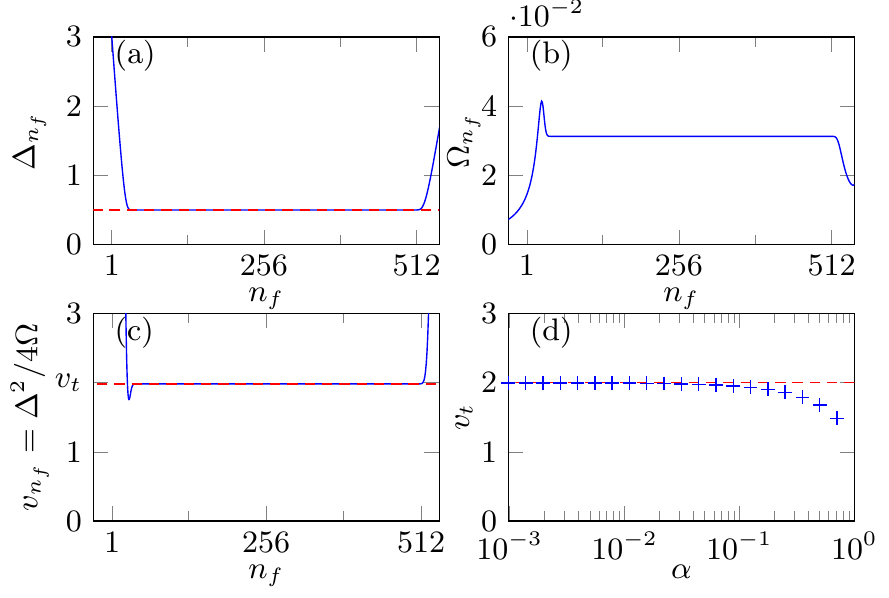} 
  \caption{Instantaneous (a) gap $\Delta(n_f)$, (b) mixing term $\Omega(n_f)$, and (c)  local value of threshold velocity  when the inhomogeneous front is traveling through the chain  for clean Ising model with $J_n =1$, $N=512$, $\alpha = 1/32$.  In (d) we show the threshold velocity in the bulk, which approaches constant for $\alpha \ll 1$.  }
  \label{fig:7}
\end{figure}

In this section,  we show the results of the analysis based on adiabatic theorem, see Sec.~\ref{sec:3}, applied to the case without disorder, $J_n = 1$. In Figure \ref{fig:7}(a-c) we show, respectively, the instantaneous gap, mixing term $\Omega(n_f)$ and, the estimation on local value of threshold velocity (from the combination of the  two using Eq.~\eqref{eq:local_vc}), for the slope of the front $\alpha=1/32$ and $N=512$. They can be directly compared with the disordered case presented in Fig.~\ref{fig:2}.

We define the threshold velocity based on the value in the bulk (i.e. when the front is inside the chain) -- we neglect here a small peak appearing in Fig.~\ref{fig:7}(c) when the front is entering the chain. The scaling of the gap in the bulk can be derived analytically,  $\Delta \simeq \sqrt{8 \alpha}$ ($\alpha \ll 1$) \cite{Dziarmaga2010a}. For small $\alpha$ we fit $\Omega \simeq \alpha$. Accordingly, as can be seen in Fig.~\ref{fig:7}(d), the vanishing of the gap with decreasing $\alpha$ is compensated by the vanishing of the mixing term that results in the saturation of $v_t$ to a constant value for $\alpha \ll 1$. This can be contrasted with the disordered case -- Fig.~\ref{fig:3}(c) -- where the threshold velocity is monotonically decreasing with (small) $\alpha$. Such a dependence is expected for models with finite critical exponent $z > 1$, see Eq.~\eqref{vc_KZ}  \cite{Dziarmaga2010b}.

The threshold velocity can be found analytically \cite{Dziarmaga2010a} as $v_t=2$ and equals  the largest velocity of quasiparticles at the critical point of the clean Ising model. We fix the factor of 4 in the denominator of Eq.~\eqref{eq:local_vc} to match the correct value of the threshold velocity in the limit of $\alpha \ll 1$. In that limit $v_t$ gives a sharp boundary between $v>v_t$ where the system gets excited when the front is traveling through the chain, and $v<v_t$, where there are no excitations appearing in the bulk.

Fig.~\ref{fig:7}(d) also indicates that for $\alpha \approx 1$ the threshold velocity is becoming smaller as the scaling prediction,  based on assumption that the front is smooth enough, $\alpha \ll 1$, does not longer hold. This is as well the limit where the first excited state is not well separated from the rest of the spectrum and the estimation of threshold velocity as presented in Sec.~\ref{sec:3} is expected to break down.

\section{Details of simulations}
The Hamiltonian in Eq.~\eqref{eq:hamiltonian} is solved in a standard way, exploiting its mapping onto a system of free fermions via the Jordan-Wigner transformation $\sigma^z_n = 1-2 c^\dagger_n c_n$;  $\sigma^x_n + i \sigma^y_n = 2 c_n \prod_{m<n}(1-2 c^\dagger _m c _m)$, where $c_n$ are fermionic annihilation operators. It is convenient to introduce Majorana fermions $a_{2n-1} = c_n + c^\dagger_n$, $a_{2n} = i (c_n-c^\dagger_n)$, which are Hermitian and satisfy anticommutation relations $\{a_m , a_n \} =  2 \delta_{m,n}$.
In this base the Hamiltonian reads:
\begin{eqnarray}
\label{eq:AH}
\hat H &=& -\sum_{n=1}^{N-1} \left( \frac{i J_n}{2} a_{2n} a_{2n+1} + h.c. \right)   \\ 
  & & - \sum_{n=1}^{N} \left( \frac{ig_n}{2} a_{2n-1} a_{2n} +h.c. \right) = \nonumber \\
  & = & \hat H_J + \hat H_g.  \nonumber
\end{eqnarray}
 For convenience we introduce  $\hat H = \vec a^\dag {\rm H} \vec a$, where $\vec a$ is a column vector consisting of operators $a_n$ and ${\rm H}$ is a $2N \times 2N$ matrix. We separate the matrix describing full Hamiltonian ${\rm H} = {\rm H}_J + {\rm H}_g$ into parts corresponding respectively to the coupling and magnetic field, which are both block diagonal, 
\begin{eqnarray*}
 & {\rm H}_g = \bigoplus_{n=1}^{N} 
\begin{pmatrix}
0   & - i g_n/2 \\
i g_n/2   & 0 \\
\end{pmatrix},\\
 & {\rm H}_J = 0 \oplus \bigoplus_{n=1}^{N-1}
\begin{pmatrix}
0 & - i J_n/2 \\
 i J_n/2  & 0 \\
\end{pmatrix} \oplus 0.
 \end{eqnarray*}

The static properties of (instantaneous) system are found by numerical diagonalisation of matrix ${\rm H}$. This amount to employing Bogoliubov transformation to a new base of Majorana fermions $\vec a = {\rm O}_0 \vec b$, where orthogonal matrix ${\rm O}_0$ brings the Hamiltonian into canonical form
$
\hat H = - i \sum_{n=1}^N  \epsilon_n b_{2n-1} b_{2n},
$
that is $ {\rm O}_0^T {\rm H}  {\rm O}_0 = \bigoplus_{n=1}^{N}
\begin{pmatrix}
0 & - i \epsilon_n/2 \\
 i \epsilon_n/2  & 0 \\
\end{pmatrix}$, with $\epsilon_n \ge 0$.  The ground state of $\hat H$ is a vacuum state annihilated by all annihilation operators in that base $(b_{2n-1} - i b_{2n})|0,n_f\rangle  =0$, and the quasiparticles energies $\epsilon_n$ are arranged in ascending order. We note that some care is needed for a system with degenerated ground state, $\epsilon_1 =0$ (within numerical precision), in which case one has to take the proper linear combination of eigenvectors of ${\rm H}$ to eigenvalue $0$, to ensure that $ {\rm O}_0$ is orthogonal.

The parity operator $\hat P = \prod_{i=1}^N \sigma^z_n = \prod_{i=1}^N (i a_{2n-1} a_{2n})$ commutes with the Hamiltonian $[\hat{H},\hat{P}] =0$, and the relevant instantaneous gap is calculated as energy of two-quasiparticles excitation, 
\begin{equation}
\label{eq:A_gap}
\Delta_{n_f} = 2(\epsilon_1 + \epsilon_2).
\end{equation}
In Fig.~\ref{fig:2} we follow the instantaneous ground state in the given parity subspace and we make sure that $\langle 0,n_f|\hat P |0,n_f\rangle$ is fixed -- and in our case equal 1. 
In case the true (numerical) ground state is a state with parity $-1$, we fix it be exciting one-quasiparticle which corresponds to $\epsilon_{1} \to  -\epsilon_{1}$, $ b_{2} \to  -b_{2}$ and  
$ [ {\rm O}_0]_{n,2} \to  -[ {\rm O}_0]_{n,2}$. The mixing term $\Omega(n_f)$ in Eq.~\eqref{eq:Omega_nf} is calculated from $\langle0,n_f|\sigma^z_n |1,n_f \rangle =\langle0,n_f| i a_{2n-1}  a_{2n} d_1^\dagger d_2^\dagger  |0,n_f \rangle$ using the Wick's theorem, where $d_n^\dagger = (b_{2n-1} + i b_{2n})/2$ creates the quasiparticle with energy $\epsilon_n$.

The time evolution in Sec.~\ref{sec:4} is simulated in the Heisenberg picture $\frac{\partial}{\partial t} a_n(t) = i [\hat H(t),a_n(t)] $.
For a free fermionic problem like in Eq.~\eqref{eq:AH} time-dependent operators can be expanded in the base of original Majorana fermions, $\vec a(t) = {\rm O}(t) \vec b$. This leads to the time-dependent Bogoliubov equations 
\begin{equation}
\frac{\partial}{\partial t} {\rm O}(t) = - 4 i {\rm H}(t) {\rm O}(t),
\end{equation}
with the initial condition ${\rm O}(t=0) = {\rm O}_0$ which corresponds to starting in the ground state of the initial Hamiltonian.
We numerically solve this differential equations by employing 4-th order time dependent Suzuki-Trotter decomposition \cite{SuzukiTrotter}, which is symplectic and allows to greatly speed up the calculations:  we split the Hamiltonian matrix ${\rm H}$ into parts corresponding to $ {\rm H}_J$ and $ {\rm H}_g$, that are block-diagonal in the original basis. This facilitates the propagation at intermediate steps  involving terms of the form $\exp(-i dt {\rm H}_{J(g)})$ that can be efficiently calculated without diagonalizing the full Hamiltonian at each time step.
Finally, the energy of the final state (here $g_f =0$) is found as $-\sum_{n=1}^{N-1} J_n \langle \sigma^x_n \sigma^x_{n+1}\rangle_{t=t_{final}} = -\sum_{n=1}^{N-1} J_n i \langle a_{2n}(t_{final}) a_{2n+1}(t_{final})\rangle$. 
 

\begin{thebibliography}{99}
\bibitem{comment_gaps} Both in homogeneous and inhomogeneous case we use $\Delta = 2(\epsilon_1 + \epsilon_2)$ (cf. \cite{Caneva2007} and Eq.~\eqref{eq:A_gap}) which is the minimal gap in the sector with given parity. Interestingly, we obtain the collapse even though in the inhomogeneous case $\epsilon_1 = 0$, which is related to part of the chain being in the ferromagnetic phase which could break the symmetry. We note, that the scaling results in \cite{Young1996,Fisher1998} were obtained for $\Delta = 2 \epsilon_1$.
\bibitem{comment_discret_alpha} We sampled over discreet values of $\alpha=2^{-s}$, with step $ds=1/2$.
\end{thebibliography}
\end{document}